\documentclass[b5paper,twoside]{jpconf}
\usepackage{graphicx}

\begin{document}
\title[A contribution of stellar flares to the GRXE]{A contribution of stellar flares to the GRXE -- based on MAXI observations --}

\author[M.Matsuoka]{M.Matsuoka$^1$, M.Sugizaki$^1$, Y.Tsuboi$^2$, 
K.Yamazaki$^2$, T.Matsumura$^2$, T.Mihara$^1$, M.Serino$^1$, 
S.Nakahira$^1$, T.Yamamoto$^1$,S.Ueno$^3$, H.Negoro$^4$ and MAXI team}

\address{$^1$Institute of Physical and Chemical Research, Wako, Saitama 351-0198}
\address{$^2$ Chuo University, Bunkyo-ku, Tokyo 112-8551}
\address{$^3$ ISAS, JAXA, Tsukuba, Ibaraki 305-8505}
\address{$^4$ Nihon University, Chiyoda-ku, Tokyo 101-8308}

\ead{matsuoka.masaru$@$riken.jp} 

\begin{abstract}
Using unbiased observations of MAXI/GSC the potential contribution of stellar 
flares and CVs to GRXE luminosity is estimated in the energy range of 2$\sim$10 keV.  
Currently, a reasonable luminosity has been obtained extrapolating  the number 
of stellar flares and that of CVs toward the Galactic ridge from those of the 
observed flares and CVs near the solar system. The ionized emission lines of 
Si to Fe are also simulated making the composite thermal spectrum which is 
based on MAXI observations of stellar flares and CVs. The present estimated 
result strongly supports a picture that the cumulative stellar flares contribute 
primarily to the GRXE in terms of the composite thermal spectrum with emission 
lines and secondary contribution is due to the thermal spectrum with high 
temperature from CVs.
\end{abstract}

\section{ Introduction to stellar flare observations by MAXI }
MAXI, the first astronomical payload on JEM-EF of ISS, has monitored all sky 
X-ray sources since August 2009, including Galactic black holes, transient 
X-ray pulsars, low mass X-ray binaries, X-ray novae, X-ray bursts, stellar 
flares, CVs, Gamma-ray bursts, numerous 
AGNs and so on \cite{mat10}. 

In this paper the detection of unexpected stellar flares has been pointed out 
in relation to a promising contribution to the GRXE \cite{tan02} \cite{rev09}, 
despite insufficient 
statistics for the detection of stellar flares.  With this in mind, a potential 
contribution to the GRXE is suggested, based on reasonable estimations of 
luminosity with emission line simulation from detected thermal spectra of 
stellar flares by MAXI/GSC. X-ray fluxes from CVs are detected as  weak 
sources, but it is possible to obtain the total luminosity of CVs in the solar 
neighborhood in order to estimate the contribution to the GRXE. 

The MAXI has two kinds of slit cameras, the GSC (Gas Slit Camera) and the SSC 
(Solid-state Slit Camera), both of which incorporate X-ray detectors 
consisting of gas proportional counters and X-ray CCDs \cite{mat09}.  
The GSC can provide  all-sky X-ray image every ISS orbit. 
The present observations of stellar flares are conducted by the GSC which 
is capable of detection with more sensitivity for stellar flares than that of any other ASMs. 

The GSC detection threshold was set at ~20 mCrab or less per ISS one orbit, hence 
the detected stellar flares were 23 for 21 months whose luminosities were 
distributed $1.6\times10^{31}$ erg/s  to  $4.8\times10^{33}$ erg/s \cite{tsu11}.  
Furthermore, 
the X-rays from Cataclysmic Variables (CVs) are generally weak 
with variable intensity and are occasionally produced with outburst.  
Currently, sources as weak as one mCrab or so are identified by making 
blinking images for one, three and seven days. Ten CVs are employed 
as effective ones to the GRXE. 
\section{ Estimation of the contribution of stellar flares to the GRXE }
Stellar flares from RS CVn's, Algol's, dMe's, young stellar 
objects and other flared stars have the common property that hot plasma 
grows up suddenly.  The wide-ranging temperature distribution is capable 
of generating the various line emissions required for the GRXE. The 
temperature of these flares ranges from \textit{kT}= a few keV to $\sim$10 keV.  All 
the data are obtained from flares of known stars near the solar system 
by unbiased observations of MAXI/GSC. 

CVs, especially magnetic CVs, generally produce thermal emission with 
higher temperature as \textit{kT}=10$\sim$40 keV \cite{ezu99}. 
The contribution of CVs to the GRXE is also estimated roughly in this paper.

\subsection{ Estimation of total luminosity of stellar flares to the GRXE }
A total luminosity from 23 stellar flares for 21 months can be estimated by
$\Sigma(luminosity[L] \times e folding time[\tau])$.  The corrected result of
the field of view as well as the live time and the energy band of 2$\sim$10 keV is 
obtained as $1.3(+0.3,-0.2)\times10^{31}$erg/s.

The stellar evolution is assumed to be equivalent to that of the neighboring
solar system.  The ratio of the stellar objects relative to the mass of the
Galactic ridge region is obtained over the mass of the neighboring solar system, using
the following two sources of data : (i) From Cox \cite{cox00}, the ratio of half 
the total mass of the Galaxy over the mass  near
the sun for the average distance, $\it d$=42 pc from observed flared stars, is 
estimated to be $6.4\times 10^6$. (ii) From Revnivtsev et al \cite{rev10}, the ratio 
of stellar mass of the Galactic ridge and bulge over stellar mass near the sun within
the average distance is estimated to be $6.9\times 10^6$.

Considering the factor of $\sim 6.5\times 10^6$ anyway, the total luminosity of
2-10 keV can be obtained for the GRXE by multiplying the value of 
$1.3\times10^{31}$erg/s as

\hspace*{20mm}  $0.85(+0.20,-0.13)\times10^{38}$  erg/s. \hspace{42mm} (1)

In this estimation the following three uncertainties are not corrected.
One uncertainty comes from flare duration time.  Here an uncertainty factor for it 
is given as $\epsilon$; e.g.,$ \epsilon \sim 0.5$.  Another uncertainty originates from
the insufficient statistics which affect the result; e.g., the GM Mus reveals an
extremely large value of $L \times \tau$, by one order of magnitude greater in
comparison to the others.  If this data were replaced tentatively by the second large
value of $L \times \tau$, the result of (1) would be $\sim 0.37 \times 10^{38}$ erg/s.
Considering these two uncertainties the result of the present estimation is expected as

\hspace*{20mm} $(0.37 \sim 0.85) \epsilon \times 10^{38}$ erg/s.  \hspace{50mm}  (2)

MAXI/GSC is considered as having a low luminosity cut-off due to instrumental capability.  In order to 
estimate entire contribution of stellar flares it is necessary to evaluate the effect of low luminosity 
flares and instrumental cut-off in low luminosity.

It is considered that the luminosity function of some stellar flares including
 solar flares has a relation of $N(L)=kL^{-\alpha}$ for the number of flares $\it{vs.} L$ 
(e.g.,\cite{car07} \cite{gue07}).  If the power law index is assumed to be $\alpha = 2$ and 
the luminosity interval of $10^{30 \sim 34} $ erg/s is effective in the GRXE, it is possible to 
evaluate an instrumental cut-off of MAXI/GSC for the low luminosity to compare the 
observed data. 

Thus obtained cut-off luminosity is around $1 \times 10^{32}$ erg/s.  
Namely a correction factor for undetectable flares is estimated to be $2\delta$, 
corresponding to a low luminosity cut-off,
where the factor, $\delta$, is introduced as remaining uncertainty; e.g., $\delta \sim 1$.
Therefore, multiplying $2\delta$ to the result value of (2), the final luminosity  for the GRXE is

\hspace*{20mm} $(0.74 \sim 1.8) \epsilon \delta \times 10^{38}$ erg/s.  \hspace{50mm}  (3)

\subsection{ Contribution of CVs and total luminosity from stellar flares and CVs }
Preliminary results of CVs are given by Matsumura et al (in preparation).  
The total luminosity of 10 CVs  observed in the solar neighborhood is effective to
the GRXE to 21-month MAXI/GSC observations.  The result of $(0.9 \sim 1.9) \times 10^{34}$ erg/s is 
obtained by estimation similar to previous section.

The average distance of CVs concerned is 282 pc.  As estimated in the section 2.1, the 
ratio of stellar objects relative to the mass of the Galaxy was obtained over the neighboring 
solar system. This ratio resulted in $\sim 2 \times 10^3$.  Finally the luminosity from stellar flares 
and CVs for GRXE is obtained as

\hspace*{20mm}  $(0.9 \sim 2.2) \epsilon \delta \times 10^{38}$ erg/s.  \hspace{52mm} (4)
\\This final value is consistent 
with the observed one, $2 \times 10^{38}$ erg/s \cite{tan02}.

\section{ Emission lines from simulated composite thermal spectrum }
MAXI/GSC cannot obtain precise thermal spectra from stellar flares. but it can determine
temperature as single temperature model.  A composite spectrum of observed stellar flares is 
created using X-ray flux $f_i$, temperature $T_i$ and e-folding time $\tau_i$ of each flare
as the equation,

\hspace*{20mm} $F(E)= \Sigma{F(E,T_i) \times f_i \times \tau_i}/\Sigma (f_i \times \tau_i) $. \hspace{30mm}  (5)
 
Figure 1 shows a composite spectrum obtained by the equation (5), where Suzaku response function
with solar abundances is employed.  Several emission lines as well as Fe 6.7 keV and
6.9 keV are seen in this figure.  

Thin thermal spectra from magnetic CVs are also expected although complicated \cite{ezu99}.  
Nevertheless, the simulated spectrum of CVs is assumed to be 
a thin thermal spectrum with \textit{kT}=10 keV.   Since the contribution of CVs to the GRXE is 
$\sim 20\%$ as obtained in previous section 2.2, 20$\%$ thermal spectrum with 
\textit{kT}=10 keV is added to the composite spectrum of stellar flares.
A final composite spectrum thus obtained from stellar flares and CVs based on 
MAXI/GSC results is simulated as shown in Figure 1, where it is noted that a contribution
lower than cut-off luminosity of stellar flares is neglected.  Now equivalent widths
of several emission lines are estimated by line fitting process 
through Suzaku analysis method.  Thus obtained equivalent widths (eV unit) are 
$\sim25$, $\sim55$,$\sim400$ and $\sim100$ for $1.8\sim2$ keV (Si), 
$2.4 \sim 2.6$ keV(S), $6.7$ keV(Fe), and $6.75 \sim7.0$ keV(Fe), respectively. 

The Fe 6.4 keV line is usually emitted by fluorescence.
In deed CVs often produce considerable strong fluorescent 6.4 keV line.  
  The 20$\%$ of
average intensity of 6.4 keV from CVs obtained by Ezuka and Ishida \cite{ezu99} is 
tentatively estimated as $\sim 30$ eV.
Furthermore, Fe fluorescent line could be produced from interstellar
gas irradiated by stellar flares.  Similar mechanism to this is proposed 
by Bond and Matsuoka \cite{bon93}
to explain Fe fluorescent line in AGN.  The equivalent width of 6.4 keV by this 
mechanism is also tentatively estimated as $\sim20$ eV.

The equivalent widths of several emission lines in this section are slightly weak in
comparison to the observed ones \cite{ebi08} \cite{yam09}.
Nevertheless, considering the present rough estimation it is 
strongly suggested that various emission lines are produced mainly
stellar flares and secondarily from CVs.

%
\begin{figure}[h]
\begin{minipage}[b]{0.5\hsize}
\includegraphics[width=17.5pc]{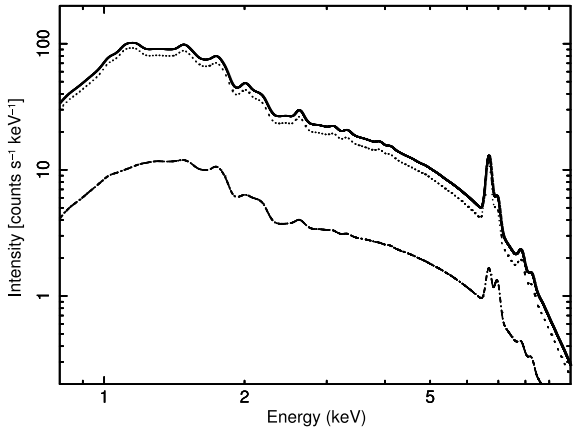}
\end{minipage}
\hspace{2pc}
\begin{minipage}[b]{0.415\hsize}
\caption{\label{label} \small A simulated composite spectrum
for the GRXE.  A chain line is a thermal spectrum with \textit{kT}=10 keV for
CVs, where the intensity level is 20\% of a composite spectrum of
stellar flares.  A dotted line is a composite spectrum of
cumulative stellar flares. A solid line is a total composite spectrum
consisting of stellar flares and CVs. In this spectrum 
Suzaku XIS response function is employed, while solar abundances are 
assumed.}
\end{minipage}
\end{figure}

\section{Conclusion}
This paper is not aimed at achieving a complete explanation of the GRXE by stellar flares and CVs, 
but it is to notice that a major of the GRXE is produced from stellar flares and the second  contribution is due to CVs.  This conclusion is based on 
the MAXI/GSC unbiased observations although statistics and detectability analysis
are not yet conclusive.  However, further MAXI/GSC observations will make those better
in future, while the assumption will be improved by progress of stellar flare physics.  
In any case the present essential conclusion would not be changed even in future 
(Matsuoka et al in preparation).



\section*{References}
\begin{thebibliography}{13} 
\bibitem{mat10} Matsuoka, M., et al. 2010, Proc. of SPIE, 7732, 77320Y-1
\bibitem{tan02} Tanaka, Y. 2002, A$\&$A, 382, 1052
\bibitem{rev09} Revnivtsev, M., et al. 2009, Nature, 458, 1142 
\bibitem{mat09} Matsuoka, M., et al. 2009, PASJ, 64, 999
\bibitem{tsu11} Tsuboi, Y., et al. 2011, Suzaku 2011 conference, SLAC, July 2011
\bibitem{ezu99} Ezuka, H. and Ishida, M., 1999, ApJ Suppl. 120, S223 
\bibitem{cox00} Cox, A.N., 2000, Allen's Astrophysical Quantities 4th ed.
\bibitem{rev10} Revnivtsev, M., et al. 2010, A$\&$A, 515, 49
\bibitem{car07} Caramazza, M., et al., 2007, A$\&$A. 471, 645
\bibitem{gue07}  Guedel, M., 2007, Rev. Solar Phys. 4, 3
\bibitem{bon93} Bond, A.I. and Matsuoka, M., 1993, MNRAS, 265, 619 
\bibitem{ebi08} Ebisawa, K., et al., 2008, PASJ, 60, S223
\bibitem{yam09} Yamauchi, S., et al., 2009, PASJ 61, S225
\end{thebibliography}
\end{document}